\documentstyle[preprint,version2,aps]{revtex}
\begin{document}

\begin{title}
Spin-Flavor Separation and Non-Fermi Liquid Behavior \\
in the Multichannel Kondo Problem: A Large N Approach
\end{title}

\author{D. L. Cox}
\begin{instit}
Department of Physics, Ohio State University, Columbus, OH 43210
\end{instit}
\moreauthors{A. E. Ruckenstein}
\begin{instit}
Serin Physics Laboratory, Rutgers University, Piscataway, NJ 08855
\end{instit}
\begin{abstract}
We consider a $SU(N)\times SU(M)$ generalization
of the multichannel single-impurity Kondo model
which we solve analytically in the limit $N\rightarrow \infty$,
$M\rightarrow\infty$, with $\gamma=M/N$ fixed.
Non-Fermi liquid behavior of the single electron Green function
and of the local spin and flavor susceptibilities occurs in both regimes,
$N\le M$ and $N > M$,
with leading critical exponents
{\em identical} to those found
in the conformal field theory solution for {\em all} $N$ and $M$ (with
$M\ge 2$).
We explain this remarkable agreement and connect it
to ``spin-flavor separation", the essential feature of
the non-Fermi-liquid fixed
point of the multichannel Kondo problem.
\end{abstract}
\pagebreak
\narrowtext

In recent years, the multichannel Kondo model,
first introduced by Nozier\`{e}s and Blandin~\cite{nozbland}, has
been the focus of intense activity.
This model involves a single local moment of spin
$S_I$ antiferromagnetically coupled to $M$ identical
conduction bands in an $N_{s}$-site system, as described by the Hamiltonian,
\begin{equation}
{\cal H} = \sum_{k\sigma\alpha} \epsilon_k c^{\dagger}_{k\sigma\alpha}
c_{k\sigma\alpha} +
({\cal J}/ N_{s}) \vec S_I \cdot [\sum_{k\sigma,k'\sigma '}
\sum_{\alpha}
c^{\dagger}_{k\sigma\alpha} \vec \tau _{\sigma \sigma '}
c_{k'\sigma '\alpha}],
\end{equation}
where $c^{\dagger}_{k\sigma\alpha}$
creates a conduction electron of (radial) wave
vector $k$,
spin $\sigma=\uparrow,\downarrow$ and channel index $\alpha$
running from $1,..,M$; the
coupling ${\cal J}>0$ is antiferromagnetic.
It is now established~\cite{betheansatz,conformal}
that in the ``overcompensated" regime,
$2S_I<M$, the screening of the local moment by the conduction electrons
drives the metal into a non-Fermi liquid critical state at $T=0$,
fully validating Nozi\`{e}res and Blandin's original arguments.

The renewed interest in this model is, in part, due to
a number of suggestions for realizations of
two-channel overcompensated behavior in explicit experimental
contexts, most notably: (i) the quadrupolar Kondo effect~\cite{coxquad}
in heavy fermion
alloys~\cite{quadexp};
(ii) non-Fermi liquid scattering
rates in narrow copper point-contacts~\cite{ralphbuh};
(iii) ``marginal" Fermi liquid normal state
properties~\cite{friends} of the high-$T_c$
materials~\cite{coxquad,emkiv}.
Moreover, it has
been argued that the two-channel Kondo model provides
a link to exotic superconductivity~\cite{conformal,coxquad,emkiv}.
Related non-Fermi liquid behavior and enhanced pairing correlations
have been recently found in
the mixed-valence regime at low temperatures in certain
extended Anderson single-impurity models motivated by the electronic
structure of the high-$T_c$ materials~\cite{mixedval} .

{}From a theoretical point of view, remarkable progress has been
made in understanding the universal properties of the multi-channel Kondo
and other impurity models through the use of conformal field theory
techniques~\cite{conformal}.
This approach provides
a clear picture of the separation of spin, channel, and charge
excitations at non-Fermi liquid fixed points and
describes the subtle recombination of these
degrees of freedom required in the Fermi liquid case.
In spite of its elegance and power this method only classifies
the {\em possible} critical behaviors without providing a
constructive route to the solution of a {\em particular} model.

In this letter we
formulate a controlled calculation method which can, in principle,
incorporate
both the complications of real materials {\em and}
the conceptual insights gained from conformal field theory.
Apart from the obvious practical application to dilute
impurity systems, this question is important in
studying the possibility for non-Fermi liquid
behavior in lattice systems, for which the conformal field theory
techniques are not immediately applicable. An explicit
route for addressing the latter problem, which appears particularly
promising,
involves constructing mean-field theories
for the lattice by solving single-impurity models embedded
in a self-consistent medium~\cite{larged}.

Below we concentrate on the $SU(N)\times SU(M)$
generalization of the multichannel Kondo Hamiltonian of Eq. (1),
where $N$ and $M$ are the degeneracies in the spin and flavor
quantum numbers, respectively. We use a functional integral
approach based on the ``slave-Boson" representation~\cite{bickers}
which explicitly separates the (local) spin and flavor excitations.
The limit $N,M \rightarrow \infty $ with $\gamma= M/N$ fixed
leads to a closed set of coupled self-consistent integral equations
which can be solved analytically
in the asymptotic low-frequency, zero temperature limit.
These are identical to
the ``Non-Crossing Approximation''(NCA) equations  of perturbation
theory~\cite{bickers}.
Although
these equations break down for sufficiently low energies in
the Fermi liquid (single channel) case~\cite{mulhart}, it is
the unique feature of our large $N,M$ treatment of
the multi-channel Kondo problem
that the NCA becomes {\em exact}.

In fact, we show that
the single electron Green function, and the local spin and
flavor susceptibilities display non-Fermi liquid behavior
with leading critical exponents
{\em identical} to those found
in the conformal field theory solution~\cite{conformal}
for {\em all} $N,M\ge 2$.
We explain this apparently surprising result
by demonstrating that the fluctuations provide
{\em no} corrections to the leading
exponents obtained from NCA {\em to all orders in $1/N,1/M$}~\cite{gan}.
Our calculation is an explicit
realization of the ``spin-charge separation" idea
emphasized by Anderson in his theory of the high-$T_c$ materials
based on the single-band Hubbard model~\cite{pwalut}.
Ultimately, we hope that the considerations presented here will be useful
in treating lattice Fermion systems
with non-Fermi liquid ground states.

Our starting point is a path integral treatment of the
generalized infinite $U$ Anderson model Hamiltonian,
\begin{equation}
H=\sum _{k,\sigma, \alpha} \epsilon _k
c^{\dagger} _{k,\sigma , \alpha } c_{k,\sigma ,\alpha } +
\epsilon _f \sum _{\sigma} f^{\dagger} _{\sigma} f_{\sigma} +
(V/\sqrt{N_{s}}) \sum _{k,\sigma ,\alpha}
[f^{\dagger} _{\sigma} b_{\bar{\alpha}} c_{k,\sigma ,\alpha} + h.c.]
\label{hamilt}
\end{equation}
where the Fermion, $f^{\dagger} _{\sigma}$, creates
a local spin excitation and the Boson, $b_{\bar{\alpha}}$, transforms
according to the conjugate representation of $SU(M)$, and
annihilates the flavor quantum number of the ``vacuum" state produced by
destroying a conduction electron.
The ``completeness" of the local states at the impurity site,
represented by the constraint,
$\sum _{\sigma} f^{\dagger} _{\sigma} f_{\sigma} +
\sum _{\bar{\alpha}} b^{\dagger} _{\bar{{\alpha}}}  b_{\bar{\alpha}} =1$,
is implemented in the usual way~\cite{bickers}, by
introducing a fictitious field $\lambda$ coupling to the constrained charges,
which is taken to $\infty$
at the end of the calculation.
In the limit
$\epsilon _f < 0, V/|\epsilon _f | \ll 1$
(\ref{hamilt}) leads to the $SU(N)\times SU(M)$ Coqblin-Schrieffer
model,
with exchange coupling ${\cal{J}} = V^2 /|\epsilon _f |$.

The most compact way of presenting our arguments
is in terms of the impurity contribution to
the partition function, $Z_{imp}$,
obtained after performing
a Gaussian integral over the
conduction electron fields. (The latter contribute
an overall multiplicative
factor, $Z_c$, of the free-electron gas partition function to
the full partition function,
$Z=Z_{c} Z_{imp}$.)
By completing squares, the hybridization term is eliminated in favor of
a spin-flavor interaction contribution,
$S_{int} = -({\tilde{V}}^2/N) \sum_{\sigma ,\bar{\alpha}}
\int d\tau \int d\tau '
f^{\dagger} _{\sigma} (\tau) f_{\sigma} (\tau ') G^0 (\tau -\tau ')
b^{\dagger} _{\bar{\alpha}} (\tau ') b_{\bar{\alpha}} (\tau)$,
where
$G^0 (\tau -\tau ') =
-\sum _{k} (\partial /\partial \tau +\epsilon _k )^{-1} /{N_s}$
is the non-interacting conduction electron Green function at the impurity site;
and, as usual, in order to obtain a nontrivial large $N,M$ limit,
we have defined a rescaled hybridization matrix element,
$\tilde{V} = \sqrt{N} V$, which should be considered of order
unity at the end of the calculation. The next step
is to introduce two composite fields, non-local in imaginary time,
$\Phi _{f\sigma} (\tau ,\tau ')$ and
$\Phi _{b\bar{\alpha}} (\tau ,\tau ')$
(with $\Phi ^{\dagger} _{f,b} (\tau ,\tau ') =\Phi _{f,b} (\tau ',\tau)$),
which decouple the
interaction term, $S_{int}$, enabling us to write,
$Z_{imp} =
\int [{\cal{D}} f][{\cal{D}} b][{\cal{D}} \Phi _f][{\cal{D}} \Phi _b]
\exp( -{\tilde{S}})$.
The effective action, $\tilde{S}$, is then given by
\begin{eqnarray}
\tilde{S}&=& \sum _{\sigma} \int _0 ^{\beta}  d\tau \int _0 ^{\beta} d\tau '
f^{\dagger} _\sigma (\tau)
[\delta (\tau -\tau' )(\frac{\partial}{\partial\tau} +\epsilon _f +\lambda )
+ \frac{\tilde{V} ^2}{N} \sum _{\bar{\alpha}}
\Phi _{b{\bar{\alpha}}} (\tau,\tau ') G^0 (\tau -\tau ')]f_{\sigma}(\tau ')
\nonumber\\
 ~&+&~
\sum _{\bar{\alpha}} \int _0 ^{\beta}  d\tau \int _0 ^{\beta} d\tau '
b^{\dagger} _{\bar{\alpha}} (\tau)
[\delta (\tau -\tau ' )(\frac{\partial}{\partial\tau} +\lambda )
- \frac{\tilde{V} ^2}{N} \sum _{\sigma} \Phi _{f\sigma} (\tau ,\tau ')
G^0 (\tau ' -\tau)]b_{\bar{\alpha}}(\tau ')\\
&-& \frac{\tilde{V} ^2}{N} \sum _{\sigma ,{\bar{\alpha}}} \int _0 ^{\beta}
d\tau \int _0 ^{\beta} d\tau'
\Phi _{f\sigma} (\tau ', \tau )
G^0 (\tau -\tau ')
\Phi _{b{\bar{\alpha}}}(\tau ,\tau ') .
\label{action}
\nonumber
\end{eqnarray}
Finally, the large $N,M$ calculation proceeds as usual by carrying out the
Gaussian integral over the local Fermions and Bosons,
$f_{\sigma}$ and $b_{\bar{\alpha}}$, to produce an effective action for
the composite fields, $\Phi _{f\sigma ,b{\bar{\alpha}}}$;
this is then evaluated by a
saddle point integration~\cite{neuberger}.
(Hereafter we drop the bar on the
channel index, $\bar{\alpha}$.)

In the $N,M=\infty$ limit with $\gamma = M/N$ fixed
the saddle point approximation becomes exact and
leads to time translationally invariant solutions,
$\Phi _{f,b} (\tau -\tau ')$.
After Fourier transforming in terms of Fermionic and Bosonic
imaginary frequencies,
$\omega _n , \nu _n$, the saddle point equations can be written as
$\Phi _f (i\omega _n ) =
[i\omega _n -\epsilon _f -\Sigma _f (i\omega _n)] ^{-1}$
and $\Phi _b (i\nu _n ) = [i\nu _n - \Pi _b (i\nu _n)] ^{-1}$. This defines
the self energies, $\Sigma _f$ and $\Pi _b$, which, on the real frequency
axis ($i\omega _n = \omega +i0 ^{+} ,i\nu _n = \omega +i0 ^{+}$), satisfy
the self-consistent
equations,
\begin{eqnarray}
\Sigma _f (\omega) &=& \frac{\gamma \tilde{\Gamma}}{\pi} \int
d\epsilon f(\epsilon ) \Phi _b (\epsilon +\omega)\\
\Pi _b (\omega)&=& \frac{\tilde {\Gamma}}{\pi} \int
d\epsilon f(\epsilon ) \Phi _f (\epsilon +\omega).
\label{selfenergies}
\end{eqnarray}
Here, $\tilde{\Gamma} =\pi \rho \tilde{V} ^2$ is the bare hybridization
width (which includes a factor of $N$ introduced by the rescaling),
and $\rho$ is the conduction electron density of states,
assumed constant in the energy range of interest. In principle, the
functions $\Sigma _f$ and $\Pi _b$ elsewhere
in the complex plane can be obtained by
the appropriate analytic continuation.
Also note that in Eq. (\ref{selfenergies})
the $\lambda \rightarrow \infty$ limit has already been
taken, and that, due to the invariance under $SU(N) \times SU(M)$ rotations
the spin and flavor indices have dropped out.

In the low-frequency, low-temperature limit,
Equations (\ref{selfenergies})
can be easily solved either by direct substitution
or by reducing them to differential
equations as done for the single channel
case by M\"{u}ller-Hartmann~\cite{mulhart}.
The relevant analysis can be found in a number of places in
the literature~\cite{bickers,mulhart}, and will not be repeated here.

The saddle-point
solutions for $\Phi _f$ and $\Phi _b$ can be written in terms of
the reduced frequency variable, $\Theta = [((1+\gamma)/\gamma )
(E_0 -\omega)/T_0 ]^{1/(1+\gamma)}$, where $E_0$ is the ground state
energy and $T_0 = D(\gamma \tilde{\Gamma} /\pi D )^{\gamma}
\exp (\pi \epsilon _f / {\tilde{\Gamma}})$ is the Kondo scale. The calculation
is most easily done for negative frequencies, $\omega < E_0$, and the
appropriate analytic continuation extends the results to $\omega > 0$.
At $T\rightarrow 0$ the resulting spin Fermion and flavor Boson
spectral functions,
$A_{f,b} (\omega)=
{\rm Im} \Phi _{f,b} (\omega-i0^+ )/\pi = A_{f,b} ^{(+)}
(\omega)\theta(\omega -E_0)$ vanish for $\omega < E_0$.
In the process of calculating physical quantities we will also need the
spectral function for occupied states ($\omega < E_0$),
defined as
$A_{f,b}^{(-)}(\omega) = \mbox{lim}_{T\to 0}
[{\rm Im} \Phi _{f,b} (\omega-i0^+ )/\pi]
\exp(\beta[E_0-\omega])$.
Close to the threshold at $E_0$, these quantities take the form
\begin{eqnarray}
A_{f}^{(+)}(\omega)&=&
\frac{1}{\pi T_0} \sin (\frac{\pi \gamma}{1+\gamma })
|\Theta|^{-\gamma} [1 + 4\frac{\gamma}{2 +\gamma} \cos(\frac{\pi \gamma}
{1+\gamma}) |\Theta| + ...]\\
A_{b} ^{(+)}(\omega)&=&
\frac{1}{\gamma \tilde{\Gamma}} \sin (\frac{\pi \gamma}{1+\gamma})
|\Theta|^{-1} [1-4\frac{W_{ch}}{1+2\gamma} \cos(\frac{\pi \gamma}{1+\gamma})
|\Theta|^{\gamma} +...]
\label{spectra}\\
A_{f}^{(-)}(\omega)&=&
\frac{\tilde{Z}}{\pi T_0}
|\Theta|^{-\gamma} [1 - 4\frac{\gamma}{2 +\gamma}
|\Theta| + ...]\\
A_{b} ^{(-)}(\omega)&=&
\frac{\tilde{Z}}{\gamma \tilde{\Gamma}}
|\Theta|^{-1} [1-4\frac{W_{ch}}{1+2\gamma}
|\Theta|^{\gamma} +...] .
\end{eqnarray}
Here, $W_{ch} = \pi T_0 /\tilde{\Gamma}$ is the weight of
channel fluctuations in the ground state, and the constant,
$\tilde{Z}$ is related to the degeneracy of the impurity ground state
per local degree of freedom (i.e., divided by $N(1+\gamma )$)~\cite{mulhart}.
Also, note the explicit breaking of particle-hole symmetry
displayed by the
positive ($A_{f,b} ^{(+)}$) and negative ($A_{f,b} ^{(-)}$)
contributions, consistent with the non-symmetric
form of (\ref{hamilt}).

The scaling dimensions of
the spin and flavor fields, $\Delta_{f} =[2(1+\gamma)]^{-1}$ and
$\Delta _b = \gamma \Delta _{f}$
can be read off
from the frequency dependence, $|E_0 -\omega|^{2\Delta _{f,b} -1}$,
in Eq. (\ref{spectra}).
It then follows that the spin, channel and physical Fermion fields,
all of which are bilinears of $f$ and $b$, have scaling dimensions,
$\Delta _{sp} = 2\Delta _f = 1/(1+\gamma)$, $\Delta _{ch}=2\Delta _b =
\gamma /(1+\gamma)$ and $\Delta _F =\Delta _f +\Delta _b =1/2$,
respectively. The resulting leading frequency dependence of the corresponding
correlation function, $|E_0 -\omega|^{2\Delta _{sp,ch,F} -1}$ is indeed
what we observe.
Below we summarize some of our explicit results.

\underline{Single Electron Green Function:}
The local electron Green function, $\cal{G} _{\sigma ,\alpha}$,
can be calculated as
a convolution of the spin and flavor propagators, leading to
a local spectral function,
$\rho_{\alpha,\mu} (\omega,0) = \mbox{Im}~{\cal{G}}_ {\sigma ,\alpha}~
{}~(\omega~-~i0 ^{+})~/\pi$, of the form,
\begin{equation}
\rho_{\alpha,\mu} (\omega,0)
\approx~
{}~ \pi / [(1~+~\gamma )^2 ~N ~\tilde{\Gamma}]~
{}~[1 + \theta(\omega) f _{+} (\tilde{\omega}) +
\theta(-\omega) f_{-} (\tilde{\omega})],
\end{equation}
with $f_{\pm} (\tilde \omega) = ( a_{\pm} |\tilde\omega|^{\Delta_{sp}} +
b_{\pm} |\tilde\omega|^{\Delta_{ch}} )$,
$a_{-}= -\left[ 4\gamma / \left( 2 + \gamma \right)\pi \right]
\sin(\pi\Delta_{ch}) B(2\Delta_{sp},\Delta_{ch})$,
$a_{+} = -\cos(\pi\Delta_{ch}) a_{-}$,
$b_{+} = -[4 W_{ch} /(1+2\gamma )\pi ] \sin(\pi\Delta_{ch})
B(2\Delta_{ch},\Delta_{sp})$
and $b_{-} = \cos(\pi\Delta_{ch}) b_{+}$.
Here $\tilde\omega=[(1+\gamma)/\gamma] (\omega/T_0 )$ and
$B(x,y)$ is the Beta function.
Note that, in the overscreened case, $M\ge N$, the leading
frequency dependence is the same as that obtained
by Affleck and Ludwig from conformal field theory~\cite{conformal}.

\underline{Resistivity:}
{}From the single-particle Green function we can obtain
the resistivity from the standard transport theory formula,
$\rho(T) \sim [\int d\epsilon ({-\partial f \over \partial\epsilon})
\tau(\epsilon,T)]^{-1} $, in terms of the scattering rate,
$\tau_{\alpha\mu}(\omega,T) ^{-1} = -2 {\rm Im}
t^{(1)}_{\alpha\mu}(\omega+i0^+,T)
=[2 \tilde{\Gamma} \rho_{\alpha\mu}(\omega,T)] /(\rho N)$,
where we have used the relation between the conduction electron $t$-matrix and
the single-electron Green function,
$t_{\alpha\mu}^{(1)}(\omega,T) = V^2 {\cal G}_{\alpha\mu}(\omega,T)$.
The resulting leading behavior,
$\rho(T)/\rho(0) \sim [1 - \alpha
(T/T_0)^{min(\Delta_{sp},\Delta_{ch})} + ... ]$ is in agreement with
that obtained by Affleck and Ludwig~\cite{conformal}. In particular,
for the special case of $N=M$,
$\Delta_{sp}=\Delta_{ch}=1/2$, we obtain a $\sqrt{T}$ correction as does
the conformal approach~\cite{conformal}.
Estimating the coefficient $\alpha$ requires a knowledge of
the temperature dependence of $\rho_{\alpha\mu}$ which
is beyond the scope of the present paper.
It is gratifying that for $N=2$
the magnitude of the spin contribution to the rate,
$[(\pi \rho) /( 2 \tau_{\alpha\mu}(0,0))] = 3 \pi ^2 /[4 (2+M) ^2 ]$,
(which is obtained after removing the $1/4$ of the total rate due to potential
scattering)
agrees with the results of
Affleck and Ludwig~\cite{conformal} to within 8\% for all $M\ge 2$.

\underline{Local Spin and Channel Flavor Dynamical Susceptibilities:}
The linear response to external fields coupling to the impurity
spin and channel quantum numbers can be easily calculated from
the bubble diagrams for the spin and channel  excitations, $f$ and $b$,
respectively. The leading and next-to-leading contributions to the
absorptive part of the
local spin and flavor susceptibilities (per spin or channel degree of
freedom), $\tilde{\chi}_{sp} '' = {\rm Im} \chi _{sp} /N$ and
$\tilde{\chi}_{ch} '' ={\rm Im} \chi _{ch} /M$, are given by:
\begin{eqnarray}
\tilde{\chi}_{sp} '' (\omega,0) &\sim &
{\gamma \Delta_{sp}^2 \sin(\pi\Delta_{ch})
\over T_0} sgn(\omega)
|\tilde\omega|^{(\Delta_{sp}-\Delta_{ch}) }
B(\Delta_{sp},\Delta_{sp}) \times \nonumber\\
&&[1-\frac{8 \gamma}{2+\gamma} \sin^2 (\frac{\pi\Delta_{ch}}{2})
\frac{B(\Delta_{sp},2\Delta_{sp})}{B(\Delta_{sp}, \Delta_{sp})}
|\tilde\omega|^{\Delta_{sp}} +...],\\
\tilde{\chi}_{ch} '' (\omega,0) &\sim &
{W_{ch} ^2 \Delta_{sp}^2 \sin(\pi\Delta_{sp})
\over T_0} sgn(\omega)
|\tilde\omega|^{(\Delta_{ch}-\Delta_{sp}) }
B(\Delta_{ch},\Delta_{ch}) \times \nonumber\\
&&[1-\frac{8 W_{ch}}{1+2\gamma} \sin^2 (\frac{\pi\Delta_{sp}}{2})
\frac{B(\Delta_{ch},2\Delta_{ch})}{B(\Delta_{ch}, \Delta_{ch})}
|\tilde\omega|^{\Delta_{ch}} +...] .
\label{susceptibilities}
\end{eqnarray}
Note that for $N=M$ (which includes the important case $N=M=2$)
both susceptibilities reduce to the general form
$\tilde{\chi} '' (\omega,0) \approx A {\rm sgn} (\omega)
[ 1- B \sqrt{{\tilde{|\omega|} / T_0}} + ...].$
This leading step function behavior in $\chi ''$ and the associated
logarithmic dependence in the real part, $\chi ' (\omega ,T) \sim
-\ln({\mbox max}\{\omega,T\}/T_0)$,
provides a possible connection with the marginal Fermi-liquid phenomenology
of the high-$T_c$ oxides~\cite{friends}. This step function behavior
was first noted
in Ref. [4(b)].
For $N > M > 1$ the real part of the spin susceptibility is
constant ($\sim 1/T_0$) and the non-Fermi-liquid behavior is dominated by
the
$|\tilde{\omega}|^{-(\Delta _{sp} -\Delta _{ch})}$
divergence of the
channel susceptibility. In the opposite limit, $N < M$,
the flavor susceptibility is constant ($\sim W_{ch} ^2 /T_0$) and a
$|\tilde{\omega}|^{-(\Delta _{ch} -\Delta _{sp})}$
divergence occurs in the spin susceptibility.
For $N>M$ the system displays two-parameter universality:
the channel fluctuations start dominating below a new energy
scale, $T_{ch} \sim T_0 W_{ch} ^{[1/(\Delta _{sp} -\Delta _{ch})]}$,
with $T_{ch} << T_0$ for $N>M$ in the Kondo limit ($W_{ch} <<1$).
We note that $T_{ch}$ evolves into the ``pathology'' temperature
below which non-Fermi liquid behavior ensues in the NCA treatment of
the one-channel Kondo model~\cite{bickers,mulhart}.

\underline{1/N Fluctuations:}
The functional integral formulation outlined above gives a natural
framework for estimating the effects
of fluctuations, and allows us to explain the remarkable fact that
the saddlepoint exponents remain unchanged to
all orders in $1/N$.
The arguments are very much
in the spirit of the argument usually made to justify the fact that
{\em perturbation theory} gives the exact exponents characterizing the
low energy behavior in conventional Fermi liquids.
More precisely, all $1/N$ fluctuations can be incorporated into
the renormalization of
interaction vertices in all diagrams ($f$ and $b$ self-energies, the
single-particle Green function as well as all susceptibilities).
We have checked explicitly to order $1/N^2$ that
these vertex renormalizations modify the
amplitudes but only give subleading
singular contributions. (For example, the sub-leading corrections
to $\Sigma _f$ and $\Sigma _b$ behave as, ${\rm Im} \delta \Sigma _f  (\omega )
\sim |\omega - E_0 |
^{3\Delta _{ch}}$
and
${\rm Im} \delta \Sigma _b (\omega) \sim |\omega -E_0 | ^{3 \Delta _{sp}}$,
respectively.)
In fact,  the saddle point gives the {\em exact}  low-energy singularities
to all orders in $1/N$,
as can be seen by
considering arbitrary order
diagrams in perturbation theory (around the saddle point).
The appropriate propagators carry
spectral functions
which diverge as
$|\omega-E_0|^{\Delta_{f,b}-1}$  [see Equations (6-9)].
As an illustration, consider a generic diagram contributing
to $\Pi _b$ which contains $L$ loops
(thus $L$ independent energy integrations),
$L$ fermion propagators,
and $L-1$ boson propagators.
The most singular contribution
behaves as $|\omega-E_0|
^{\zeta_b(L)}$ where $\zeta_b(L) =
L  +L(\Delta_f -1) + (L-1)(\Delta_b-1)= 1-\Delta_b$
since unitarity of the scattering amplitude
requires $\Delta_f+\Delta_b=1$. This is indeed the behavior found at the
saddle point.
Similar arguments apply to $\Sigma_f$,
the one particle
Green's function and the spin and channel
susceptibilities.
It is natural to speculate that,
in all systems with
spin-charge separated, non-Fermi liquid ground states,
the correct low energy behavior can be
obtained on the basis of an appropriate (self-consistent) perturbation theory
involving the ``separated" spin and charge degrees of freedom.
In analogy with conventional Fermi liquids, such perturbative arguments
should be valid even when
no obvious small parameters are available,
provided no phase transition occurs to a Fermi liquid state
through the ``binding" of
spin and charge.

\underline{Crossover Effects:}
A related issue concerns the crossover to the Fermi liquid solution which
becomes the correct ground state in the presence of a spin or
channel symmetry breaking field,
respectively $H_{sp}$ or $H_{ch}$.
Our saddle point calculation leads to a crossover
away from multi-channel behavior below respective scales,
$T_{sp}\sim H_{sp}^{1+1/\gamma}, T_{ch}\sim H_{ch}^{1+\gamma}$.
The corresponding crossover
exponents,
$\phi_{sp} =1+1/\gamma , \phi_{ch} =1+\gamma$,
are precisely those obtained from conformal field theory~\cite{conformal}.
However, the Fermi liquid
behavior which sets in well below the crossover scale $T_{sp(ch)}$
is outside the scope of the NCA.
This can be traced back to the fact that the Kondo screening characteristic of
the Fermi liquid fixed point involves the formation of a singlet bound state
between a conduction electron and the local spin excitation,
$f_{\sigma}$~\cite{wolfle};
the residue of the bound state pole plays the role of the slave-boson
mean field amplitude in the conventional large $N$ approach to
the Kondo problem.
In the multichannel problem this Fermi liquid saddle point
becomes possible only in the
presence of a channel symmetry breaking field.

Above we have shown that the limit $N\rightarrow \infty , M\rightarrow \infty$
with $M/N=\gamma$ fixed allows us to obtain the {\em exact} low energy
behavior of the multichannel single impurity Kondo problem.
It suggests the possibility of a phenomenological approach
to the multichannel Kondo problem and other impurity models with
non-Fermi liquid ground states based on spin-flavor (or spin-charge) separated
degrees of freedom. This can be regarded as the logical extension
of Nozieres' classic discussion of the single channel
Fermi liquid case to non-Fermi liquid situations.
A detailed analysis of
fluctuations about the multichannel saddle point, including the
crossover to the Fermi liquid case will be described elsewhere.
Moreover, this formulation leads to a natural extension
to the lattice through the large-$D$ treatment of
correlated systems~\cite{larged}.

We wish to acknowledge useful discussions
with I. Affleck, S. Barle, M. Jarell, A. Ludwig, M. Makivic, Q. Si, C. M. Varma
and
P. W\"{o}lfle.
A.E.R. is especially greatful to G. Kotliar for a
stimulating conversation concerning the role
of fluctuations about the NCA saddle point.
This research was supported in part
by a grant from the U.S. Department of Energy,
Office of Basic Energy Sciences, Division of Materials Research (D.L.C.), and
by
ONR Grant \# N00014-92-J-1378 and a Sloan Foundation Fellowship (A.E.R).

\pagebreak

{\bf Figure} Phase diagram in the $N,M$ plane.  The
NCA is strictly controlled in the large $M,N$ limit for $M\ge 2$,
but gives the exact
critical exponents {\em for all} $N,M\ge 2$.
The universality class is the same for all
lines of fixed slope $\gamma=M/N$.

\end{document}